\documentclass[aps,epsfig,prl,preprint,superscriptaddress]{revtex4}
\usepackage{amsmath}
\usepackage{graphicx}

\newcommand{\bsigma}{\mbox{\boldmath $\sigma$}}

\begin{document}

\title{Valley-antisymmetric potential in graphene under dynamical deformation}

\author{Ken-ichi Sasaki}
\email{sasaki.kenichi@lab.ntt.co.jp}
\affiliation{NTT Basic Research Laboratories, NTT Corporation,
3-1 Morinosato Wakamiya, Atsugi, Kanagawa 243-0198, Japan}

\author{Yasuhiro Tokura}
\affiliation{NTT Basic Research Laboratories, NTT Corporation,
3-1 Morinosato Wakamiya, Atsugi, Kanagawa 243-0198, Japan}
\affiliation{Graduate School of Pure and Applied Sciences,
University of Tsukuba, Tsukuba, Ibaraki 305-8571, Japan}

\author{Hideki Gotoh}
\affiliation{NTT Basic Research Laboratories, NTT Corporation,
3-1 Morinosato Wakamiya, Atsugi, Kanagawa 243-0198, Japan}

\date{\today}

\begin{abstract}
 When graphene is deformed in a dynamical manner, a time-dependent
 potential is induced for the electrons.
 The potential is antisymmetric with respect to valleys, and some
 straightforward applications are found for Raman spectroscopy.
 We show that a valley-antisymmetric potential broadens Raman $D$ band
 but does not affect $2D$ band, which is already observed by recent
 experiments.
 The space derivative of the valley antisymmetric potential gives
 a force field that accelerates intervalley phonons, while it corresponds to the
 longitudinal component of the previously discussed pseudoelectric field
 acting on the electrons.
 Effects of a pseudoelectric field on the electron is quite difficult to
 observe due to the valley-antisymmetric coupling constant, on the other hand, 
 such obstacle is absent for intervalley phonons with $A_{1g}$
 symmetry that constitute the $D$ and $2D$ bands.
\end{abstract}

\maketitle

A lattice structure with exact periodicity, namely equal bond lengths
for every bond species, does not exist in nature. 
Lattice deformation is inevitable even in a simple planar structure
consisting of a single atom species, namely graphene.~\cite{novoselov05,zhang05}
Graphene research has already clarified that 
lattice deformation can be expressed by a gauge
field.~\cite{gonzalez92,kane97,lammert00,sasaki05,jackiw07,sasaki08ptps,Vozmediano2010}
A deformation-induced gauge field has been employed in predicting notable phenomena. 
For example, Landau levels, which are usually formed by applying a
magnetic field to graphene, are also constructed by applying strain to
graphene. 
The novel Landau levels were predicted in terms of a static
strain-induced pseudomagnetic field, which is defined by the rotation of
the gauge field by analogy with a magnetic field.~\cite{guinea10,Levy2010}
There is a crucial difference between magnetic and pseudomagnetic fields.
Namely, the coupling constant of pseudomagnetic field has opposite
signs for the electrons at different valleys (valley anti-symmetric),
whereas the coupling constant of a magnetic field is the same for the
valleys (valley symmetric).

Recently, some attempts have been made to understand the effects of
dynamical (time-dependent) lattice deformation in terms of a
pseudoelectric field, which is defined by the time derivative of the
gauge field by analogy with an electric
field.~\cite{Vozmediano2010,Vaezi2013,Trif2013}
A pseudoelectric field as well as a normal electric field can accelerate
the electrons and induce current.
However, since the coupling constant is valley anti-symmetric, 
the currents at the K and K' valleys flow in the opposite
directions and they cancel out. 
As a result of the cancellation effect, 
a pseudoelectric field does not cause a net electric current, in general. 
While several authors try to extract a net electric current from a
time-dependent lattice deformation,
such attempt is based on unrealistic assumptions.~\cite{Vaezi2013}

In this paper, we show that the cancellation effect of a pseudoelectric field
which originated from the valley-antisymmetric coupling constant,
is absent for intervalley phonons.
The phonons are observed in the Raman spectrum of graphene,
and the effects of a pseudoelectric field are already visible from
recent experiments of Raman spectroscopy. 
The point of our theoretical analysis is that 
a valley-antisymmetric potential is induced for the electrons when graphene
is deformed in a dynamical manner, and 
the spatial derivative of the potential gives a force field for the
phonons, which turns out to be the longitudinal component of the
previously discussed pseudoelectric field acting on the electrons.
In other words, we have extended the idea of a pseudoelectric field to
include the phonons, by introducing the concept of a
valley-antisymmetric potential acting on the electron.
The force field on the phonons is free from the cancellation effect so
that the effect of a pseudoelectric field is easy to observe. 

\begin{figure}[htbp]
 \begin{center}
  \includegraphics[scale=0.4]{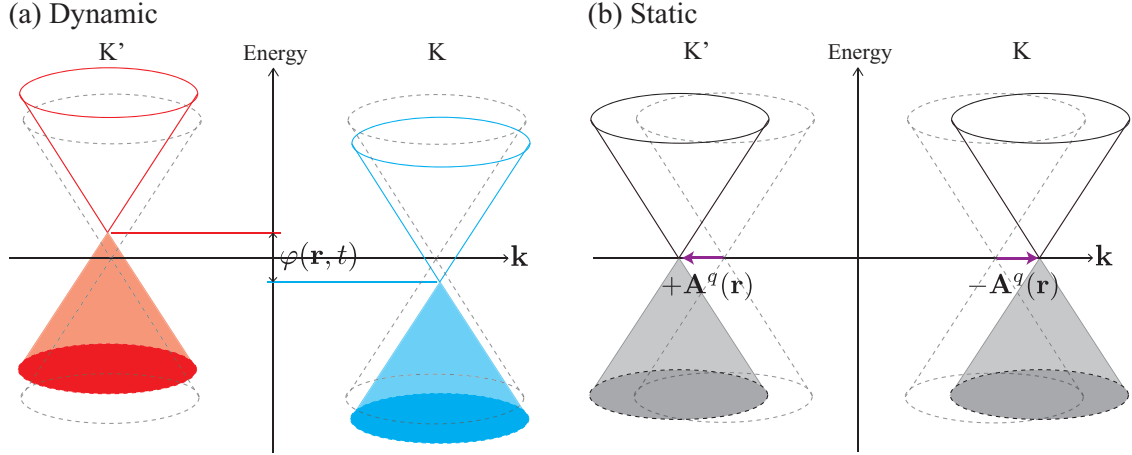}
 \end{center}
 \caption{
 (a) A transient valley-antisymmetric potential is induced when
 lattice deformation is time dependent.
 (b) When the deformation becomes static, the potential disappears, while the
 positions of the K and K' points are slightly
 shifted compared with those in the absence of deformation.
 }
 \label{fig:Dirac-cone}
\end{figure}

Before providing an exact formulation of a valley-antisymmetric potential,
we outline the main result here in such a way that readers will easily
be able to grasp the essential points.
When graphene is deformed in a time dependent manner,
the electrons are subject to a potential.
It is valley antisymmetric, that is, 
the sign of the potential is different
for the K and K' valleys as shown in Fig.~\ref{fig:Dirac-cone}(a).
The potential is time dependent and disappears when the deformation
becomes static as shown in Fig.~\ref{fig:Dirac-cone}(b).
Meanwhile, static deformation causes a shift of the wavevector between
the K and K' points, which is permanent.

\begin{figure}[htbp]
 \begin{center}
  \includegraphics[scale=0.7]{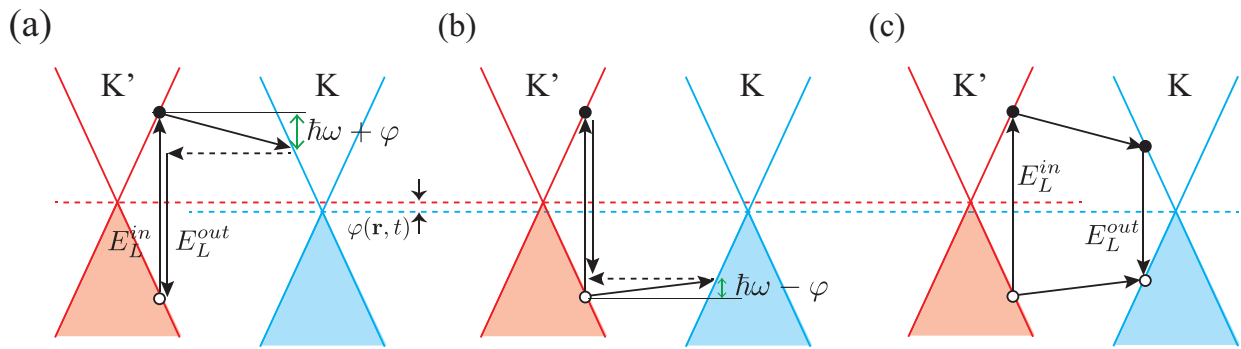}
 \end{center}
 \caption{
 The origin of the $D$ band broadening in the presence of a
 valley-antisymmetric potential $\varphi$.
 (a) When a photo-excited electron experiences an intervalley scattering
 by emitting an $A_{1g}$ phonon with frequency $\omega$, the Raman
 shift $E_L^{in}-E_L^{out}$ is given by $\hbar \omega + \varphi$.
 (b) When a hole emits an $A_{1g}$ phonon, $E_L^{in}-E_L^{out}$ is given by $\hbar \omega - \varphi$.
 (c) There is no spectrum change for the $2D$ Raman band because of the
 cancellation: $E_L^{in}-E_L^{out}=(\hbar\omega + \varphi)+ (\hbar\omega - \varphi)=2\hbar \omega$.
 }
 \label{fig:Dshift}
\end{figure}

A straightforward application of a valley-antisymmetric potential is found
with respect to the defect-induced $D$ band observed in the Raman spectrum of
graphene.~\cite{tuinstra70,Cancado2004,Sasaki2013}
The Raman process starts when an electron-hole pair is
created in the Dirac cones by absorbing a laser light, as shown in
Fig.~\ref{fig:Dshift}(a) and~\ref{fig:Dshift}(b).
A photo-excited electron or hole can change its valley 
by emitting an intervalley phonon, which constitutes the $D$ band.
Elastic intervalley scattering (represented by dashed arrows) caused by certain
defects, such as the armchair graphene edge,~\cite{Cancado2004,gupta09,Sasaki2013} 
can compensate for the changes in the valleys.
Then the electron recombines with the hole as the result of a scattered light
emission, which completes the Raman process.~\cite{malard09}
Let $\omega$ and $\varphi$ be the frequency of the phonon and
the potential difference, respectively.
As shown in Fig.~\ref{fig:Dshift}(a),
the potential difference $\varphi$ contributes to the Raman shift, 
which is the energy difference between the incident and scattered light
as $E_L^{in}-E_L^{out}=\hbar \omega + \varphi$.
On the other hand, when the hole emits an intervalley phonon, 
the energy difference is $\hbar \omega - \varphi$ as shown in 
Fig.~\ref{fig:Dshift}(b).
Because the processes shown in Fig.~\ref{fig:Dshift}(a)
and~\ref{fig:Dshift}(b) occur with equal probability, the $D$ band broadens in the Raman spectrum,
provided that $\varphi$ is distributed in the spot of the laser light.
The broadening is interpreted as the emission of a combination mode of the
intervalley phonon and acoustic phonon, as shown later at Eq.~(\ref{eq:fnew}).
The $2D$ Raman band, which consists of two intervalley phonons, does not
broaden even when $\varphi\ne 0$ because of the cancellation:
$(\hbar\omega + \varphi)+ (\hbar\omega - \varphi)=2\hbar \omega$,
as shown in Fig.~\ref{fig:Dshift}(c).
The effect of a valley antisymmetric potential on the $2D$ band is in sharp
contrast to that on the $D$ band.

Such $D$ and $2D$ Raman band behavior was observed in a recent Raman
spectroscopy experiment.~\cite{Suzuki2013}
In the experiment, H$_2$O molecules are considered to be inserted
between the graphene layer and substrate.
When H$_2$O molecules are irradiated by a laser light,
the interaction between H$_2$O molecules and graphene induces a
time-dependent lattice deformation, which we identify as the origin of $\varphi$.
The interaction between H$_2$O molecules and graphene is also discussed
by Mitoma {\it et al.}~\cite{Mitoma2013}
The authors observed an enhancement of the $D$ band intensity for
graphene on a water-rich substrate and explained it in terms of the
photo-oxidation of graphene.
For a strong laser light, the deformation can be significantly large and 
graphene may be further damaged by such dynamical deformation.

\begin{figure}[htbp]
 \begin{center}
  \includegraphics[scale=0.5]{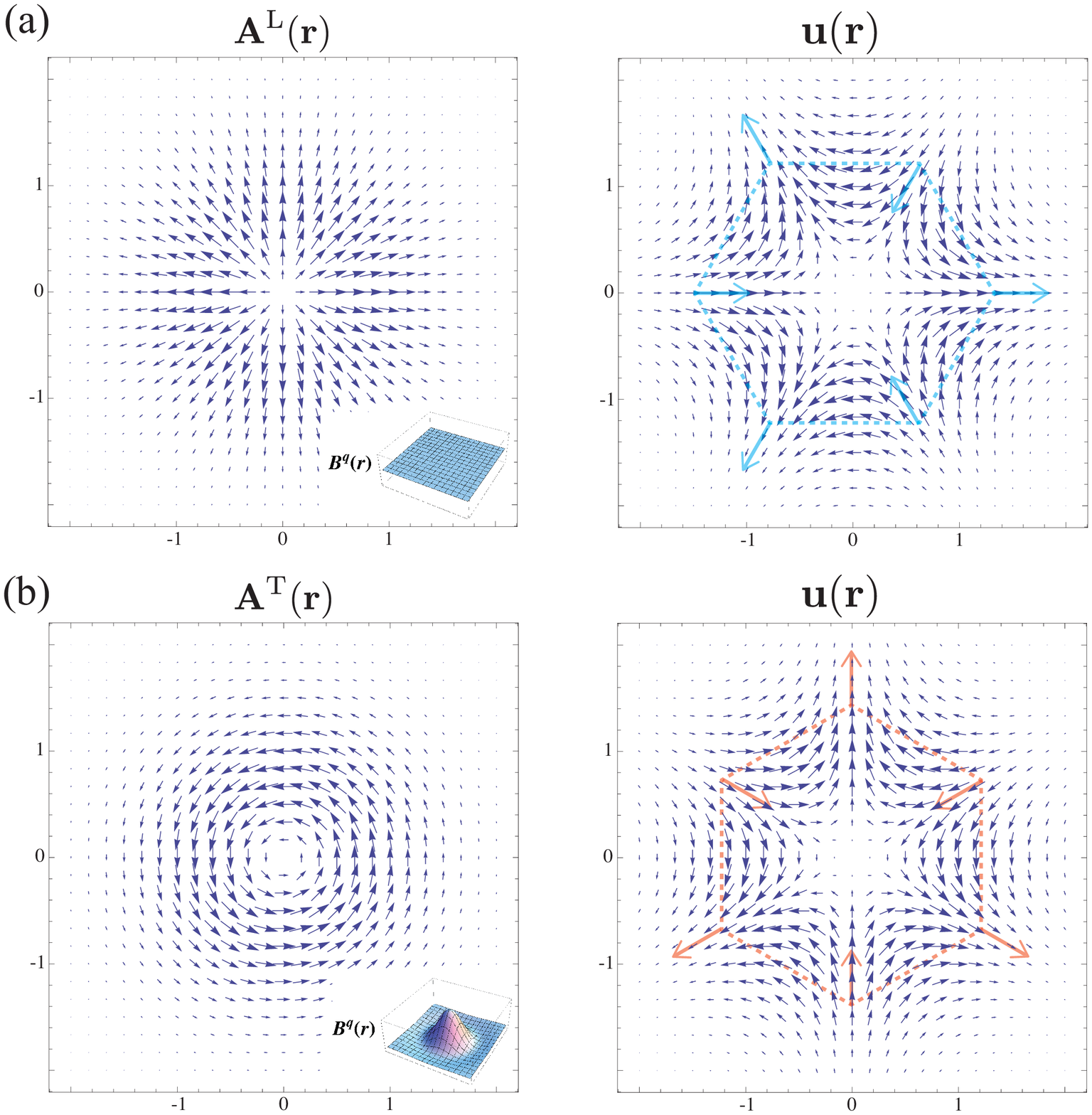}
 \end{center}
 \caption{
 (a,left) Plot of longitudinal deformation-induced gauge
 field ${\bf A}^{\rm L}({\bf r},t) = \nabla \Phi({\bf r})$ with $\Phi({\bf
 r})\propto -e^{-({\bf r}/\ell)^2}$. 
 Note that when we assume that ${\bf A}^{q}({\bf r},t) = \nabla
 \Phi({\bf r}) \cos(\omega t)$, the vector field also represents
 the pseudoelectric field ${\bf E}^q({\bf r},t)$.
 The inset shows the pseudomagnetic field $B^{q}({\bf r})$ which is zero
 for ${\bf A}^{\rm L}({\bf r},t)$. 
 (a,right) The corresponding displacement vector field ${\bf u}({\bf r})$. 
 The arrows represent the direction of the field at the points.
 (b,left) Plot of transverse gauge field ${\bf A}^{\rm T}({\bf r},t) = (\partial_y,-\partial_x) \Phi({\bf r})$.
 The inset shows the plot of pseudomagnetic field $B^{q}({\bf r})\ne 0$. 
 (b,right) The corresponding displacement vector field ${\bf u}({\bf r})$.
 }
 \label{fig:gauge}
\end{figure}

A valley antisymmetric potential and a shift of the K and K' points are both
described in a unified fashion by the formulation of a deformation-induced gauge field.
When a C-C bond expands or shrinks in the graphene plane,
the hopping integral diverges from the equilibrium value $-\gamma$.~\cite{porezag95}
Thus, lattice deformation is defined by a change in the hopping
integral, which depends on the position of a carbon atom ${\bf r}$, 
the bond direction $a$ ($=1,2,3$), and time $t$, 
as $-\gamma + \delta \gamma_a({\bf r},t)$.
In the absence of deformation, 
the low-energy dynamics of the $\pi$-electrons is
governed by a massless Dirac equation.~\cite{novoselov05,zhang05}
Lattice deformation modifies the massless Dirac equation into a Dirac
equation that includes a deformation-induced gauge field
${\bf A}^q({\bf r},t)$ and mass $m({\bf r},t)$ as~\cite{sasaki08ptps} 
\begin{align}
 i\hbar \frac{\partial}{\partial t}
 \begin{pmatrix}
  \Psi_{\text K}({\bf r},t) \cr \Psi_{\text K'}({\bf r},t)
 \end{pmatrix}
 =
 \begin{pmatrix}
  v \bsigma \cdot (\hat{\bf p}+{\bf A}^{q}({\bf r},t)) & m({\bf r},t) \sigma_x \cr
  m({\bf r},t)^* \sigma_x & v\bsigma' \cdot (\hat{\bf p}-{\bf A}^{q}({\bf r},t))
 \end{pmatrix}
 \begin{pmatrix}
  \Psi_{\text K}({\bf r},t) \cr \Psi_{\text K'}({\bf r},t)
 \end{pmatrix}.
 \label{eq:Deq}
\end{align}
In this equation, $v\sim 10^6$ m/s is the velocity of the electron,
$\hat{\bf p}=-i\hbar \nabla$ is the momentum operator,
$\bsigma=(\sigma_x,\sigma_y)$ and $\bsigma'=(-\sigma_x,\sigma_y)$
are the Pauli matrices that operate on the two-component wavefunction 
for each valley, $\Psi_{\text K}$ or $\Psi_{\text K'}$. 
The mass is induced when the lattice deformation has a short distance
periodicity comparable to the lattice constant.
In the lattice vibrational motions in graphene,
such deformation appears as an intervalley phonon with $A_{1g}$ symmetry.~\cite{tuinstra70,malard09,Cancado2004,Sasaki2013}
Let ${\bf k}_F+{\bf q}$ be the wavevector of an $A_{1g}$ phonon, 
where ${\bf k}_F$ denotes the wavevector of the K point, 
and $\omega$ is the frequency, it can be shown that 
$m({\bf r},t)=m_0 e^{i{\bf q}\cdot {\bf r}}\cos(\omega t)$, 
where $m_0$ is a constant determined by the electron-phonon
coupling $g \sim 6$ eV.
Meanwhile, the field ${\bf A}^q({\bf r},t)=(A_x^q({\bf r},t),A_y^q({\bf r},t))$ 
is calculated from the displacement vector of the acoustic phonons
${\bf u}({\bf r},t)=(u_x({\bf r},t),u_y({\bf r},t))$ by 
$vA_x^q({\bf r},t)=g( -\partial_x u_x({\bf r},t) + \partial_y u_y({\bf r},t) )/2$ and 
$vA_y^q({\bf r},t)=g(\partial_y u_x({\bf r},t) + \partial_x u_y({\bf
r},t))/2$ (See Sec.I of Supplementary Materials for details).
Some examples of static gauge field ${\bf A}^q({\bf r})$ are shown in Fig.~\ref{fig:gauge}.

The crucial step in clarifying the effect of the time dependence of
lattice deformation is to decompose ${\bf A}^{q}({\bf r},t)$ into
divergenceless (transverse mode) and rotationless (longitudinal mode) 
fields that satisfy $\nabla \cdot {\bf A}^{\rm T}({\bf r},t) = 0$ and 
$\nabla \times {\bf A}^{\rm L}({\bf r},t)  = 0$,
respectively, as~\cite{sasaki05}
\begin{align}
 {\bf A}^{q}({\bf r},t) = {\bf A}^{\rm L}({\bf r},t) + {\bf A}^{\rm
 T}({\bf r},t).
 \label{eq:decomp}
\end{align}
This decomposition is always possible because of the Helmholtz's theorem (See Sec.II of Supplementary Materials).
The field ${\bf A}^{\rm T}({\bf r},t)$ induces a non-zero pseudomagnetic
field, ${\bf B}^{q}({\bf r},t) \equiv \nabla \times {\bf A}^{\rm T}({\bf r},t)$,
while the field ${\bf A}^{\rm L}({\bf r},t)$ induces zero pseudomagnetic
field. 
Thus, ${\bf A}^{\rm L}({\bf r},t)$ can be written as the space
derivative of a scalar function $\Phi({\bf r},t)$ as
${\bf A}^{\rm L}({\bf r},t) = \nabla \Phi({\bf r},t)$.
By putting $\Psi_{\text K}({\bf r},t) = e^{-i\Phi({\bf r},t)/\hbar}\psi_{\text K}({\bf r},t)$ 
and $\Psi_{\text K'}({\bf r},t) = e^{+i\Phi({\bf r},t)/\hbar}\psi_{\text K'}({\bf r},t)$ into
Eq.~(\ref{eq:Deq}), the Dirac equation is 
written in terms of the new wavefunction 
$(\psi_{\text K}({\bf r},t),\psi_{\text K'}({\bf r},t))$ as
\begin{align}
 i\hbar \frac{\partial}{\partial t}
 \begin{pmatrix}
  \psi_{\text K}({\bf r},t) \cr \psi_{\text K'}({\bf r},t)
 \end{pmatrix}
 = 
 \begin{pmatrix}
  v \bsigma \cdot (\hat{\bf p}+{\bf A}^{\rm T}({\bf r},t) ) - \dot{\Phi}({\bf
  r},t)\sigma_0 & m({\bf r},t)e^{+2i\frac{\Phi({\bf r},t)}{\hbar}} \sigma_x \cr 
  m({\bf r},t)^* e^{-2i\frac{\Phi({\bf r},t)}{\hbar}}\sigma_x &
  v\bsigma' \cdot (\hat{\bf p}-{\bf A}^{\rm T}({\bf r},t) ) + \dot{\Phi}({\bf r},t)\sigma_0
 \end{pmatrix} 
 \begin{pmatrix}
  \psi_{\text K}({\bf r},t) \cr \psi_{\text K'}({\bf r},t)
 \end{pmatrix}.
 \label{eq:Deq_new}
\end{align}
In this representation, the effects of the time dependence are apparent.
First, $\dot{\Phi}({\bf r},t)$ 
($\equiv \partial \Phi({\bf r},t)/\partial t$) appears as a potential for
the electrons in the K and K' valleys.
Note that the signs of the potentials are reversed for the two
valleys.~\footnote{
The acoustic phonons are generally area-changing deformations, for which a
time-dependent potential appears at the diagonal components 
of Eqs.~(\ref{eq:Deq}) and (\ref{eq:Deq_new}).
Thus, it may be appropriate to call it valley-asymmetric potential rather than
valley-antisymmetric potential. See Supplement for details.
} 
As a result, there is a potential difference between the valleys,
\begin{align}
 \varphi({\bf r},t) \equiv 2\dot{\Phi}({\bf r},t).
 \label{eq:varphi}
\end{align}

Secondly, $\Phi({\bf r},t)$ appears at the phase of the off-diagonal
terms.
Thus, for the $A_{1g}$ phonons, the off-diagonal parts are given by
$m_0 e^{\pm i\left( {\bf q}\cdot {\bf r}+2\Phi({\bf r},t)/\hbar
\right)}\cos(\omega t)\sigma_x$.
The wavevector and frequency of the modified $A_{1g}$ phonon,
which are defined as the space and time derivatives of the phase, are 
\begin{align}
 & {\bf q} \to {\bf q}+2\frac{\nabla \Phi({\bf r},t)}{\hbar}, 
 \label{eq:qderit} \\
 & \hbar \omega \to \hbar \omega \pm \varphi({\bf r},t). 
 \label{eq:fnew}
\end{align}
This confirms the broadening of the $D$ Raman band, which we discussed earlier.
The $A_{1g}$ phonon is softened and hardened 
in a dynamical manner, which is sharp contrast to 
the static strain-induced
softening of an $A_{1g}$ phonon.~\cite{Marianetti2010,Si2012}
For static deformation, we have 
${\bf q} \to {\bf q}+2\nabla \Phi({\bf r})/\hbar$
and $\varphi({\bf r},t)=0$,
which show that only the wavevector of the $A_{1g}$ phonon is changed,
and the wavevector shift is due to the translation of the Dirac cones
as shown in Fig.~\ref{fig:Dirac-cone}(b), namely, 
${\bf k}_F \to {\bf k}_F - \nabla \Phi({\bf r})/\hbar$.
Specifically, the shift of ${\bf q}$ may result in a shift of the peak
position of the $D$ band because $\omega$ is linearly dependent on
$|{\bf q}|$.~\cite{vidano81}
Note that a shift of the peak position of the $D$ band is independent of
the broadening, while the broadening of the $D$ band may be related with
the lifetime of the $A_{1g}$ phonon. 
However, the $D$ band broadening observed in
Ref.~\onlinecite{Suzuki2013} cannot be attributed only to the lifetime
of the $A_{1g}$ phonon. 
Because, if the lifetime of the $A_{1g}$ phonon is the main cause of the
broadening of the $D$ band, the $2D$ band that consists of two $A_{1g}$
phonons should broaden too, which is not observed in the experiment.
Furthermore, the lifetime of the $A_{1g}$ phonon calculated is too long
to explain the broadening. 
The long lifetime of the $A_{1g}$ phonon results from the fact that the
decay of the $A_{1g}$ phonon into an electron-hole pair is 
suppressed by the kinematics.~\cite{sasaki12_migration}

Thirdly, 
$\nabla \dot{\Phi}({\bf r},t)= \dot{\bf A}^{\rm L}({\bf r},t)$ 
accelerates the electrons (induces electric currents) while the
directions of the force fields are opposite for the K and K' valleys.
In addition to this, ${\bf A}^{\rm T}({\bf r},t)$ 
causes a shift of the wavevector between the K and K' points as shown in
Fig.~\ref{fig:Dirac-cone}(b).
When $\dot{\bf A}^{\rm T}({\bf r},t)=0$, the shift is static, while
when $\dot{\bf A}^{\rm T}({\bf r},t)\ne 0$ the shift becomes dynamical
and
accelerates the electrons (induces electric currents) at the K and K'
points.
Since we have defined a pseudomagnetic field as 
${\bf B}^q({\bf r},t) = \nabla \times {\bf A}^{q}({\bf r},t)$
by analogy with a magnetic field ${\bf B}({\bf r},t) = \nabla \times {\bf A}({\bf r},t)$,
it is reasonable to define a pseudoelectric field 
as ${\bf E}^q({\bf r},t)= -\dot{\bf A}^{q}({\bf r},t)$
by analogy with an electric field 
${\bf E}({\bf r},t) = - \dot{\bf A}({\bf r},t)$.
Thus, ${\bf E}^{q}({\bf r},t)$
is referred to as the
pseudoelectric field in literature.~\cite{Vozmediano2010}
Because these currents flow in the opposite directions,
they are subject to the cancellation and do not cause a net electric
current.~\footnote{
However, when graphene is connected to a charge reservoir,
the electrons can be transferred between the reservoir and graphene 
to establish equilibrium.
If the transfer is instantaneous and smooth, 
a non-zero valley-antisymmetric potential results in a change in the Fermi
energies in the K and K' valleys.
In this case, electric currents induced by $\dot{\bf A}^{\rm T}({\bf
r},t)$ at the K and K' points do not cancel, and a net electric current
appears in a time-dependent manner.
}

Though the pseudoelectric field has been considered as a force
field for the electron, it also causes a force field for the $A_{1g}$ phonon.
To show this, we first note that Eq.~(\ref{eq:qderit}) can be expressed
in the differential form,
\begin{align}
 \hbar \dot{\bf q}=2 \dot{\bf A}^{\rm L}({\bf r},t).
 \label{eq:eqofphonon}
\end{align}
The right-hand side is reminiscent of the electric field that
accelerates the electron, which appears in the equation of motion;
\begin{align}
 \hbar \dot{\bf k}=-e {\bf E}({\bf r},t).
 \label{eq:eom}
\end{align}
By employing a gauge transformation,
the electric field is expressed as the time derivative of the gauge field
(vector potential) 
${\bf A}({\bf r},t)$ as ${\bf E}({\bf r},t) = - \dot{\bf A}({\bf r},t)$ 
without losing generality.
This makes it easy to see that there is a strong similarity between
Eqs.~(\ref{eq:eqofphonon}) and (\ref{eq:eom}).
Namely, the left-hand side of the equations is the time derivative of
the momentum and right-hand side is the time derivative of the gauge field.
Using the decomposition of Eq.~(\ref{eq:decomp}), ${\bf E}^{\rm
L/T}({\bf r},t) = - \dot{\bf A}^{\rm L/T}({\bf r},t)$, 
Eq.~(\ref{eq:eqofphonon}) is written as 
$\hbar \dot{\bf q}=-2{\bf E}^{\rm L}({\bf r},t)$, which shows that 
a longitudinal component of the pseudoelectric field is a force field
that accelerates the $A_{1g}$ phonon. 
This definition is consistent with the natural thought that the force is
the space derivative of the potential: 
${\bf E}^{\rm L}({\bf r},t)= - \nabla \varphi({\bf r},t)$.
From symmetry point of view, 
${\bf E}^{\rm L}({\bf r},t)$ or $\varphi({\bf r},t)$ is the measure of valley asymmetry, while 
${\bf B}^q({\bf r},t)$ is the origin of the asymmetry between
sublattices.~\cite{sasaki08ptps} 
Since the pseudo-electromagnetic fields cover all degrees of freedom of
the electrons in graphene, we can expect there to be a set of partial
differential equations for ${\bf E}^q({\bf r},t)$ and ${\bf B}^q({\bf r},t)$, 
similar to Maxwell's equations for electromagnetic field.
It is, however, beyond the scope of this paper.

Generally, to know $\varphi({\bf r},t)$,
we need to solve a dynamical $\Phi({\bf r},t)$ equation.
Such an equation is obtained from the potential energy functional for
$\Phi({\bf r},t)$ (See Sec.III of Supplementary Materials);
\begin{align}
 E[\Phi] = \left(\frac{2v}{g}\right)^2 \int d^2{\bf r} \left\{ \frac{\rho}{2} \dot{\Phi}^2 + \frac{\mu}{2}
 (\nabla \Phi)^2 + F[\Phi] \right\},
\end{align}
where $\mu$ is Lam\'e coefficient and $\rho$ is the density of matter.
The last term $F[\Phi]$ is a potential energy that 
depends on the lattice deformation characterized by 
annealing temperature and adsorbates.
Moreover, $F[\Phi]$ is also dependent on ${\bf B}^q({\bf r},t)$ (or ${\bf A}^{\rm
T}({\bf r},t)$), and $\varphi({\bf r},t)$ is difficult to identify.
Here, let us assume that $E[\Phi]$ has a local minimum at $\langle \Phi({\bf r}) \rangle \ne 0$
and that the time evolution is given by 
$\Phi({\bf r},t) = \langle \Phi({\bf r}) \rangle + \delta \Phi({\bf
r})\cos(\omega_0 t)$, 
where the last term represents a fluctuation.
From Eq.~(\ref{eq:varphi}), 
$\varphi({\bf r},t)= 2\omega_0 \delta\Phi({\bf r})\sin(\omega_0 t)$ 
where $\omega_0$ is the typical frequency of the fluctuation. 
Note that Eq.~(\ref{eq:Deq_new}) is invariant when $\Phi({\bf r},t)$ is
replaced with $\Phi({\bf r},t)+\pi \hbar n$, where $n$
is an integer, which suggests the existence of degenerate ground states.
If the electronic system has a periodicity in time $T=\pi/\omega_0$, 
$2\delta \Phi({\bf r})$ must be a multiple of $\pi \hbar$.
Thus, the amplitude of $\varphi({\bf r},t)$ is given by $\pi \hbar \omega_0 n$.

To conclude, dynamical lattice deformation
induces a time-dependent potential for the electrons 
that is antisymmetric in valleys. 
The valley-antisymmetric potential appears when dynamical fluctuation
occurs around a static deformation.
The space derivative of the valley-antisymmetric potential,
that accelerates the $A_{1g}$ phonon, was identified as the longitudinal
component of a pseudoelectric field.
In the presence of a pseudoelectric field, 
the energy of the $A_{1g}$ phonon undergoes softening and
hardening in a dynamical manner, which is reflected in the behavior of
the Raman $D$ and $2D$ bands.
A further experimental validation of the effect of a pseudoelectric
field on these Raman bands is of prime importance because 
the pseudoelectric field together with the previously discussed
pseudomagnetic field cover all the degrees of freedom of the electrons
in graphene, which are the valleys and sublattices.

\bibliographystyle{apsrev4-1}
%

%

\end{document}